\documentclass[amssymb,aps,floats,preprint,prb]{revtex4}
\usepackage{color,graphicx,pstricks,subfigure,bm, amsmath}
\usepackage{overpic}
\draft


\makeatletter
\long\def\@makecaption#1#2{%
  \par
  \vskip\abovecaptionskip
  \begingroup
   \small\rmfamily
   \sbox\@tempboxa{%
    \let\\\heading@cr
    #1$\,\bm|$ #2%
   }%
   \@ifdim{\wd\@tempboxa >\hsize}{%
    \begingroup
     \samepage
     \flushing
     \let\footnote\@footnotemark@gobble
     #1$\,\bm|$ #2\par
    \endgroup
   }{%
     \global \@minipagefalse
     \hb@xt@\hsize{\hfil\unhbox\@tempboxa\hfil}%
   }%
  \endgroup
  \vskip\belowcaptionskip
}%
\renewcommand\@biblabel[1]{{#1}}
\makeatother

\newcommand{\critical}[1]{$#1_{\mathrm{c}}$}
\newcommand{\Tc}{\critical{T} {}}

\newcommand{\ybco}{YBa$_2$Cu$_3$O$_{7-\delta}$}

\newcommand{\beq}{\begin{equation}}
\newcommand{\bea}{\begin{eqnarray}}
\newcommand{\eeq}{\end{equation}} 
\newcommand{\eea}{\end{eqnarray}}

\begin{document}

\title{\,
\vspace*{4 mm}
$\mbox{\boldmath$h/e$}$-Periodicity in Superconducting Loops
\vspace{2cm}}

\author{{}
F.~Loder$^1$, A.\,P.~Kampf$^1$, T.~Kopp$^{1,*}$, J.~Mannhart$^1$,  
C.\,W.~Schneider$^1$, and Y.\,S.~Barash$^2$
\vspace{0,5cm}}

\affiliation{$^1$Center for Electronic Correlations and Magnetism, Institute 
of Physics, \\ University of Augsburg,
D-86135 Augsburg, Germany
\\
$^2$Institute of Solid State Physics, Russian Academy of Sciences,
Chernogolovka, Moscow District, 142432 Russia 
\\
\\
\\
\hbox{$^*$\rm To whom the correspondence should be addressed: 
thilo.kopp@physik.uni-augsburg.de}
}
 
\vspace{2cm}
\date{\today}\vspace{1.5cm}


\vspace{1cm}

\maketitle

{\bf 
The magnetic flux periodicity of superconducting loops as well as flux quantization itself are a 
manifestation of macroscopic quantum phenomena with far reaching implications.They provide 
the key to the understanding of many fundamental properties of superconductors and are the 
basis for most bulk and device applications of these materials.
In superconducting rings the electrical current has been known to periodically respond 
to a magnetic flux with a periodicity of $\bm{h/2e}$. Here, the ratio of Planck's constant and 
the elementary charge defines  the magnetic flux quantum $\bm{h/e}$. The well-known 
$\bm{h/2e}$ periodicity  is viewed to be a hallmark for electronic pairing in superconductors and is 
considered  evidence for the existence of Cooper pairs. Here we show that in contrast to this 
long-term  belief, rings of many superconductor bear an $\bm{h/e}$  periodicity. These superconductors 
include the  high-$\bm{T_c}$ cuprates, Sr$_2$RuO$_4$, the heavy-fermion superconductors, 
as well as all other unconventional superconductors with nodes in the energy gap functions, 
and s-wave superconductors with small gaps or states in the gap. As we show, the 50-year-old
Bardeen--Cooper--Schrieffer theory of superconductivity implies that for multiply connected 
paths of such superconductors the ground-state energies and consequently also the supercurrents 
are generically $\bm{h/e}$ periodic.  The origin of this periodicity is a magnetic-field driven reconstruction 
of the condensate and a concomitant Doppler-shifted energy spectrum. The robust, flux induced 
reconstruction of the condensate will be an important aspect to understand nanoscale properties 
of unconventional superconductors. 
}

Currents of electrons moving on multiply connected paths are modulated by an applied magnetic flux with a period of $h/e$~$^($\cite{Olario}$^)$, as predicted by Aharonov and Bohm \cite{AB} . 
In superconducting rings the order parameter responds also periodically to a 
magnetic flux, as Fritz London recognized when he analyzed the implications of 
a single-valued superconducting wave function \cite{London}; different 
condensate states, which differ by integer flux quanta, are related by a gauge transformation. 
London concluded that the flux periodicity in superconducting rings is 
$h/e$~$^($\cite{London}$^)$. He missed, however, a class of supercurrent carrying wave 
functions,  which 
were identified years later \cite{Byers,Brenig, Onsager}, and allowed to explain the 
experimentally observed $h/2e$ flux quantization~\cite{Doll,Deaver}. 
Indeed, according to the Bardeen-Cooper-Schrieffer (BCS) theory of 
superconductivity \cite{BCS} the electronic condensate is formed by Cooper 
pairs, which carry twice the elementary charge. However, fundamentally it is 
not just the pairing motivated substitution of $e$ by $2e$, from which the periodicity in 
$h/2e$ originates, but rather the subtle requirement of the degeneracy in 
energy~\cite{Byers, Brenig, Onsager} of the two distinct classes of 
supercurrent carrying states.

The original flux trapping experiments  \cite{Doll, Deaver}, which proved 
the $h/2e$ flux quantization in superconductors,
%
as well as the later experiments \cite{Parks,Abrikosov,Essmann} were considered 
a manifestation of the formation of Cooper 
pairs in the then known conventional superconductors. The discovery that 
magnetic flux changes the magnetization of \ybco {\,} rings with a periodicity 
of $h/2e$ was similarly argued to provide the evidence for Cooper pairs also in
high-temperature superconductors \cite{Gough}.
%

Does, vice versa, the existence of 
Cooper pairs or the $h/2e$ flux quantization necessarily imply an $h/2e$ 
periodicity of the energy or the current in superconducting loops? In 
fact, the $h/2e$ periodicity requires that multiply connected superconductors 
threaded by a flux $n\, h/2e$ are degenerate in energy for different integers 
$n$. In superconducting $s$-wave rings or hollow cylinders with inner diameter 
$d$ this degeneracy occurs if $d\gg\xi$, where $\xi$ is the coherence length \cite{Byers, Brenig, Onsager}. In the opposite regime 
$d\lesssim\xi$ the discrete quantum nature of the electronic states in the ring matters and the energies at half-integer and integer flux 
quanta are generally different; correspondingly the superconducting behavior is only $h/e$-periodic (see 
Fig. 8-8 in Ref.~\onlinecite{Schrieffer}). This behavior should be observable,
possibly in Al rings with $d<\xi=1.6\,\mu$m.

The oscillation period of energy and currents in superconducting rings is 
therefore not always $h/2e$. In fact, as we report here, the BCS-theory 
strictly predicts that for rings of superconductors with nodes in their gap 
functions, such as the high-\Tc cuprates, Sr$_2$RuO$_4$, or the heavy-fermion 
superconductors, the ground-state energy is generically $h/e$ periodic. For 
all these superconductors, the states that yield the BCS-condensate state
also include current-carrying states with energies close to the Fermi energy 
$E_F$. As a result of the magnetic-field driven change of occupation of these 
states and the concomitant reconstruction of the condensate, the 
superconducting rings develop an $h/e$ periodicity of the supercurrent.

The flux periodicity in mesoscopic loops of $d$-wave superconductors is 
contained in the solution of the Bogoliubov--de Gennes (BdG) 
equations \cite{deGennes} for the pairing Hamiltonian:
$$
{\cal H}=-t\sum_{\langle ij \rangle,\sigma}
            e^{i \varphi_{ij}}c^\dagger_{i\sigma} c_{j\sigma}
            +\sum_{\langle ij\rangle}\left[ \Delta^*_{ij}
             c_{j\downarrow}c_{i\uparrow}+ \Delta_{ji}c_{i\uparrow}^\dag c_{j\downarrow}^\dag\right].
\label{BCS}
$$
The operators $c_{j\sigma}$ ($c^\dagger_{j\sigma}$) annihilate 
(create) an electron on lattice site $j$ with spin $\sigma=
\uparrow,\downarrow$; $t$ is the hopping matrix element between nearest 
neighbor sites, $\varphi_{ij}=2\pi e/h\int^j_i {\bf A}({\bf r})\cdot{\rm d}
{\bf r}$ is the Peierls phase factor, and ${\bf A}$ is the vector potential 
of the magnetic field. The order 
parameter of the superconducting state $\Delta_{ij}$, defined on the links between neighboring sites with phase factors appropriate for $d$-wave symmetry.

Figure~1 
displays the probability density of the wave function for a state with energy 
close to $E_F$ on a square loop, whose edges are oriented parallel to the [100] and 
[010]-directions, respectively. 
\begin{figure}[t!]
\centering
\includegraphics[width=9cm]{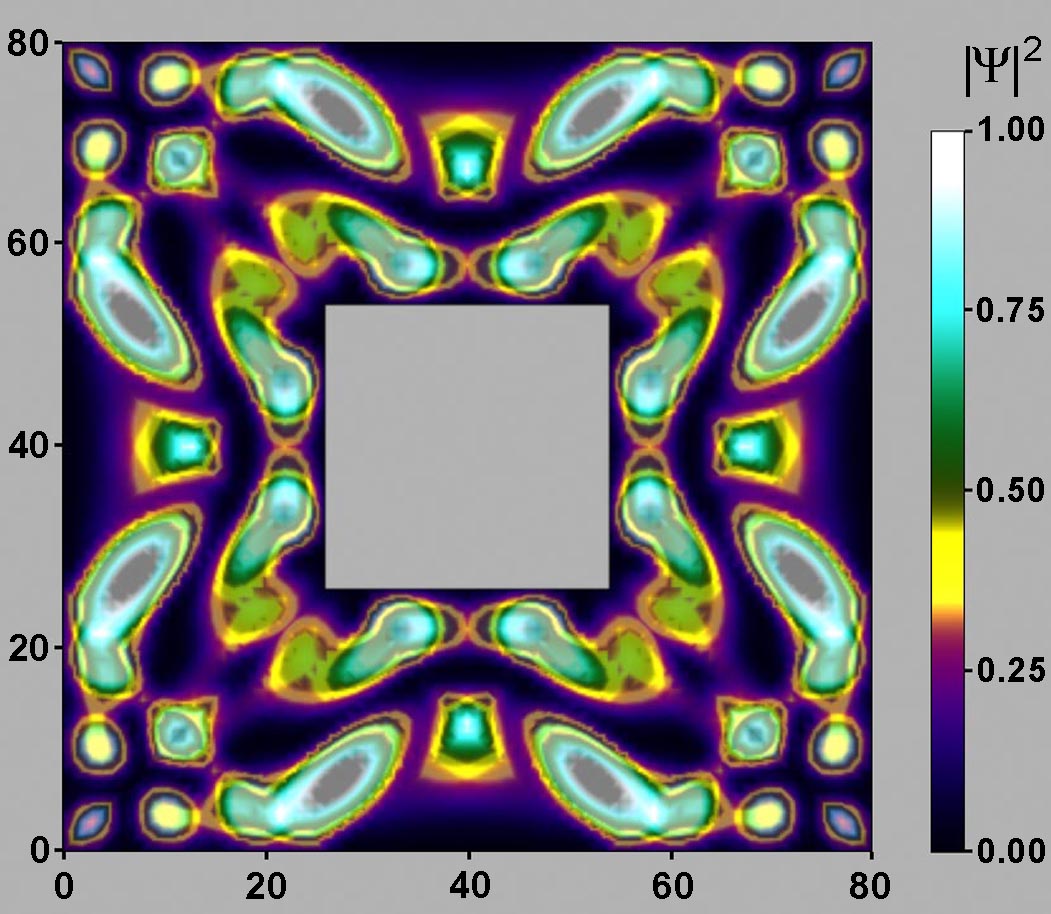}
\caption{\setlength{\baselineskip}{5mm}
{\bf Real-space representation of a square loop with a typical electronic probability density $\bm{|\Psi|^2}$
of a single state in the condensate.} The figure displays an eigenstate of the $d$-wave pairing Hamiltonian, 
calculated for a  square-loop with 80$\times$80 lattice sites with a pairing interaction of 
$0.3t$. The hole in the center has a size of 28$\times$28 unit cells.
To enhance the contrast of the complex pattern, the special color
code shown on the right is used and the discrete lattice points are smoothly 
interpolated.}
\label{Fig1}
\end{figure}
The $d$-wave loop eigenstates are obviously far more complex than
the angular momentum eigenstates of a one-dimensional circular ring 
(cf.\ Ref.~\onlinecite{Schrieffer}), and the current flow in this loop can 
only be evaluated numerically. Nevertheless, also a qualitative discussion 
allows insight into the underlying physics. 

To assess the global quantities, viz. energy and current, the evolution of the 
eigenenergies with magnetic flux has to be calculated. The eigenstates with 
energies below $E_F$ form the ground-state condensate (Fig.~2). Only flux values $\Phi$ between 0 and $h/2e$ are discussed, because all quantities are 
either symmetric or antisymmetric with respect to flux reversal $\Phi\rightarrow -\Phi$. Two clearly distinct 
regimes are found: the flux intervals between 0 and $h/4e$ and 
from $h/4e$ to $h/2e$.

Up to $\Phi=h/4e$ the supercurrent $J$ generates a magnetic 
field which tends to reduce the applied field.
This is achieved by a continuous shift of the eigenenergies in the 
condensate.  At $\Phi=0$, pairs of states with opposite circulation compensate 
their respective currents, thus $J=0$. The well separated states at $\Phi=0$ 
in Fig.~2 are the states in the vicinity of the nodes of the mesoscopic 
$d$-wave superconductor. At energies further away from 
$E_F$, the state density is higher; these are the states near the maximum 
energy gap that provide most of the condensation energy. For $\Phi>0$, the 
energy of the states with orbital magnetic moment anti-parallel (parallel) to 
the magnetic field is increased (decreased). Correspondingly the supercurrent, 
which is carried by these states, depends on the details of level crossings 
and avoidings. The main contribution to 
the supercurrent arises from the occupied levels closest to $E_F$, because the 
contributions from the lower-lying states tend to cancel in adjacent 
pairs. 

\begin{figure}[t!]
\centering
\includegraphics[width=11cm]{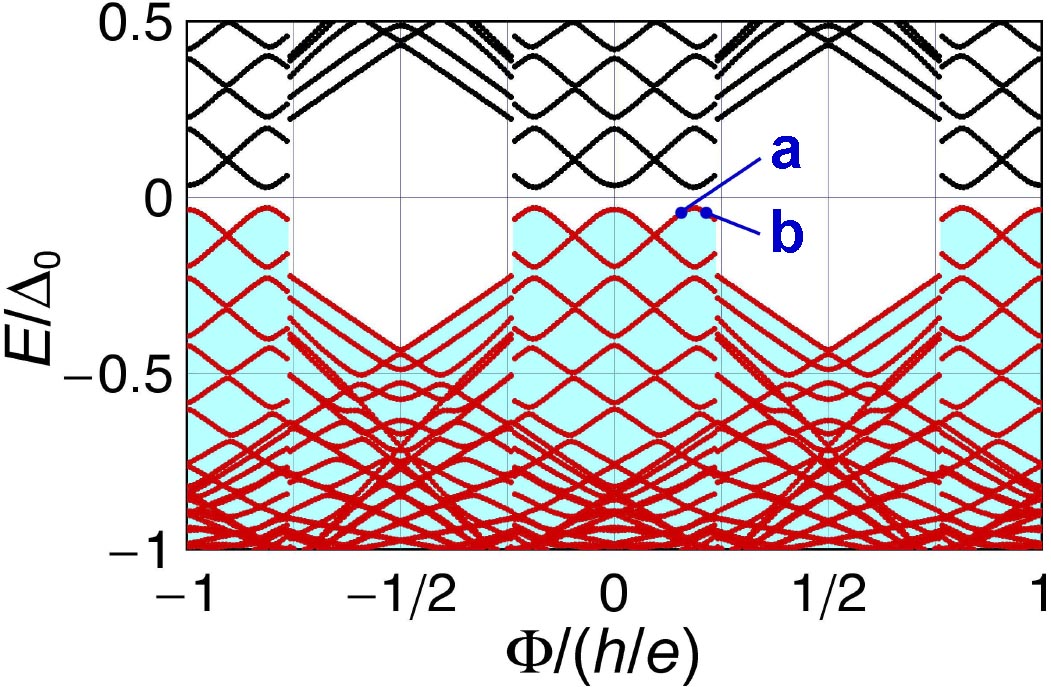}
\caption{\setlength{\baselineskip}{5mm}
{\bf Energy spectrum of the $\bm d$-wave BCS model.} The eigenenergies in the gap region are shown for a square 
40$\times$40 loop with a hole of 14$\times$14 unit cells and 
pair interaction $0.3t$ as a function of flux $\Phi$ (in units of $h/e$). The 
energies are given in units of the superconducting order parameter $\Delta_0$ 
at $\Phi=0$ ($\Delta_0\approx0.22t$). The superconducting condensate consists of the states below 
$E_F=0$ (red lines). Reconstruction of the condensate takes place near 
$\Phi=\pm(2n+1)h/4e$, where the eigenenergies jump abruptly.}
\label{Fig2}
\end{figure}

As the highest occupied state shifts with increasing flux to lower energies,
the current in the square loop first increases for small $\Phi$ (Fig.~3), 
then decreases, when the highest occupied level with an orbital moment 
opposite to the applied magnetic field starts to dominate. With increasing 
flux this state approaches $E_F$. For $s$-wave rings this ``Doppler shift 
energy'' (cf.~Ref.~\onlinecite{deGennes}) corresponds to the critical value of 
the superfluid velocity, for which the indirect energy gap closes. For $d$-wave loops, the order parameter 
is protected by the numerous states that form the ``lobes'' of the $d$-wave gap 
parameter.

\begin{figure}[t!]
\centering
\includegraphics[width=11cm]{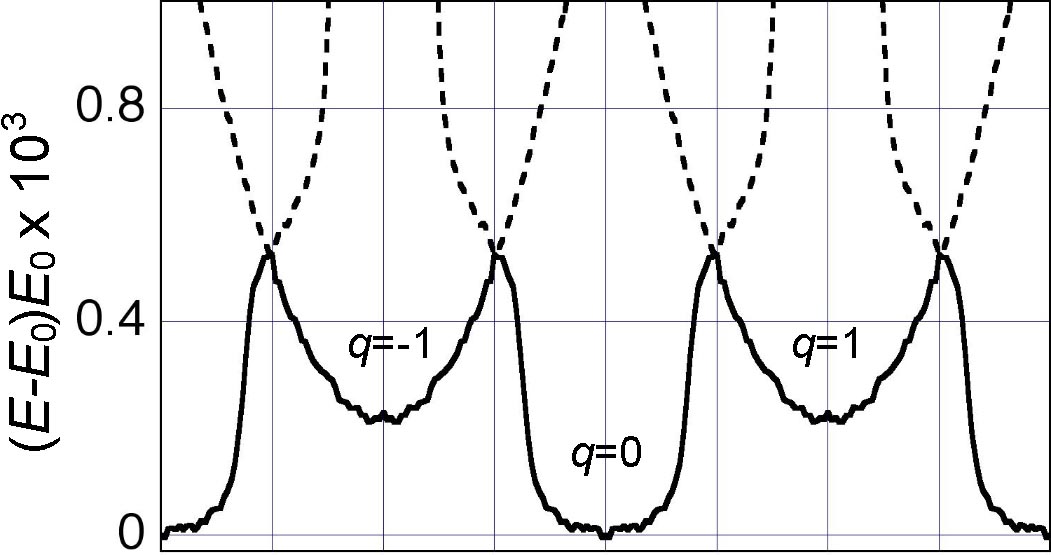}\\[1mm]
\includegraphics[width=11cm]{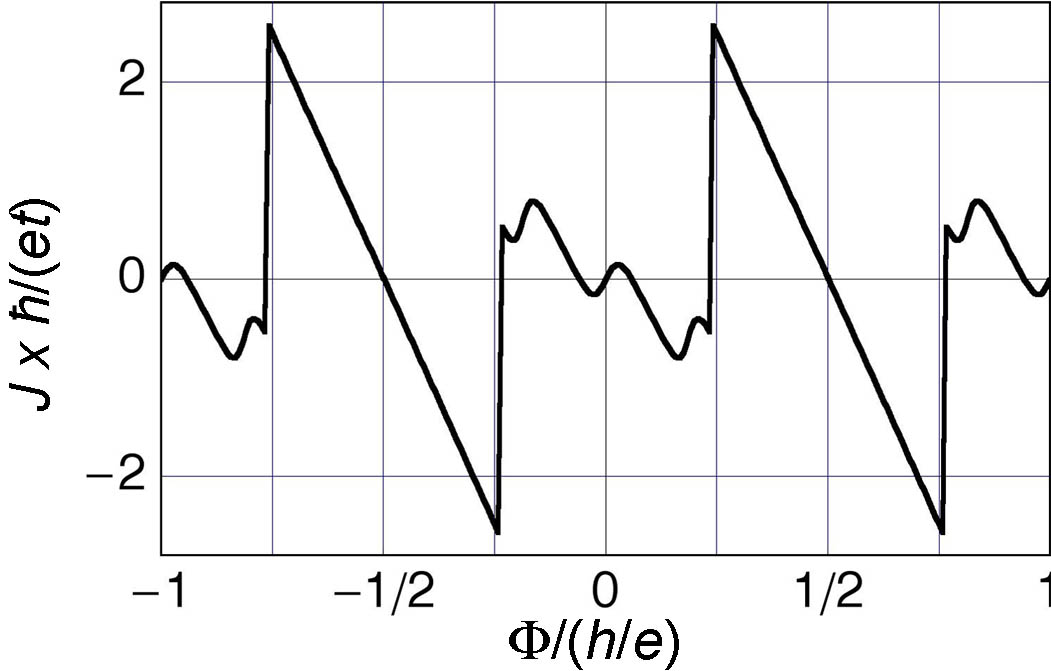}
\caption{\setlength{\baselineskip}{5mm}
{\bf Flux dependence of energy and current.} Total energy $(E(\Phi)-E(0))/E(0)$ ({\bf a}) and total circulating current $J$ ({\bf b}) 
for a square 40$\times$40 loop with a hole of 14$\times$14 unit cells and 
pair interaction $0.3\,t$ as a function of flux $\Phi$ in units of $h/e$. $J$ is 
given in units of $et/\hbar=6\times10^{-5}A$ for a typical choice of $t=250$~meV. 
There is a clear difference between condensate states with an even and an odd
winding number $q$ of the order parameter, reflected e.g. in the deformation of 
the $q=0$-parabola. The overall $\Phi$-periodicity for $E$ and $J$ is $h/e$.} 
\label{Fig3}
\end{figure}

For $d$-wave loops and rings with other unconventional order 
parameter symmetries, the states in the vicinity of the 
nodes evolve with increasing flux as in small gap $s$-wave rings. 
They do not necessarily cross $E_F$ (Fig.~2) due the hybridization of the respective
states above and below $E_F$. Nevertheless, a state with one direction of
current is replaced by a state of opposite direction (Fig.~4). 
The current carrying states of the condensate are thereby 
continuously changing near the extrapolated crossing points. As a consequence
the energy ``parabola'' centered at zero flux is different from the ground-state 
energy parabola centered at $\Phi=h/2e$ (Fig.~3). The deviation from a 
parabolic shape near zero flux is due to the evolution of the near-nodal states; 
the vertical offset of the energy minima at $\Phi=nh/e$ results mostly from the flux 
dependence of the states near the maximum value of the 
anisotropic gap. 

For a flux value near $h/4e$ the condensate reconstructs. The superconducting 
state beyond $h/4e$ belongs to the class of wavefunctions introduced by Byers 
and Yang~\cite{Byers}. Remarkably, in the flux interval 
from $h/4e$ to $h/2e$, a full energy gap exists also for $d$-wave superconductors 
(Fig.~2). Here the circulating current
enhances the magnetic field; the paramagnetic moment of the 
current is parallel to the field. The resulting energy gain is responsible for
the field-induced energy gap. This reconstruction of the condensate is the origin of 
the $h/e$ periodicity in energy and current.
Intriguingly, for superconductors with unconventional order 
parameter symmetries also larger loops ($d\gg \xi$) are $h/e$ periodic. 

\begin{figure}[t!]
\centering
\begin{overpic}
[width=8.4cm]{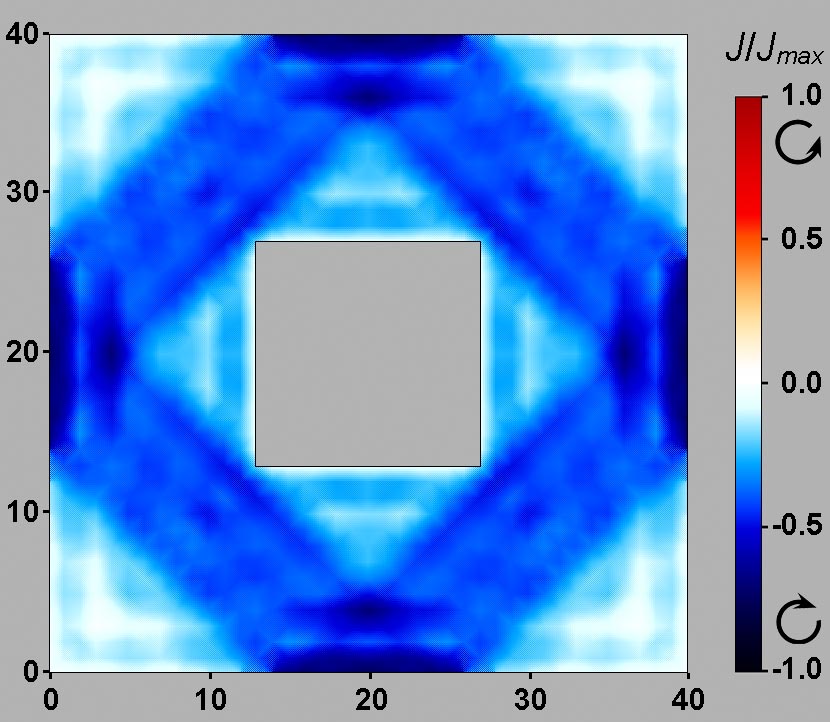}
\put(-5,82){\large\bf{a}}
\end{overpic}
\\[2mm]
\begin{overpic}
[width=8.4cm]{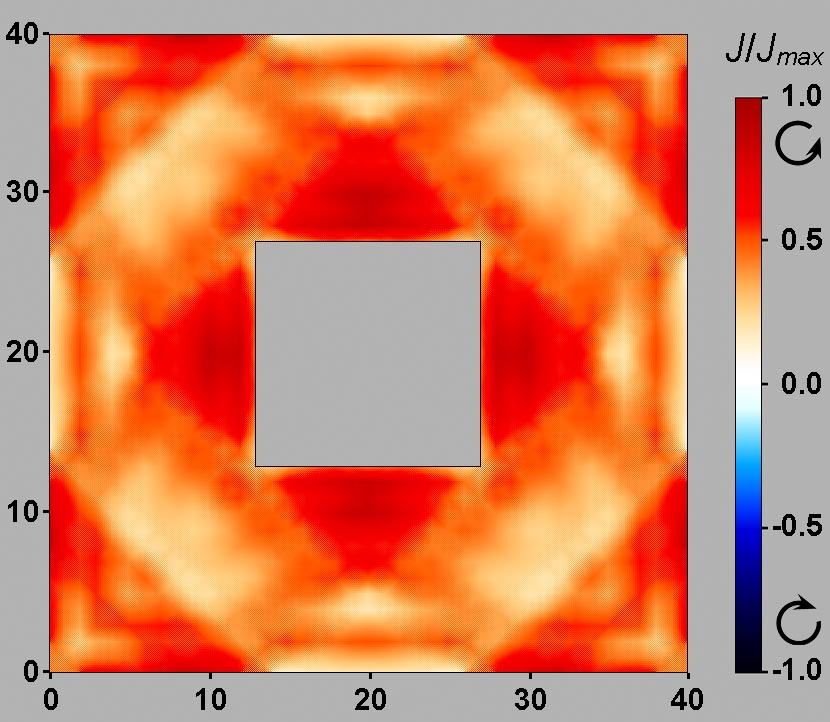}
\put(-5,82){\large\bf{b}}
\end{overpic}
\caption{\setlength{\baselineskip}{5mm}
{\bf Current distribution in a square loop of 40$\times$40 lattice sites.}
The current expectation value of the occupied state closest to $E_F=0$ is shown for flux 
$\Phi=0.17\, h/e$ (top panel, the state is marked with `{\bf a}' in Fig. 2) and for $\Phi=0.21\, h/e$ (bottom
panel, marked with `{\bf b}' in Fig. 2). The color encodes the projection of the current onto
a square path around the loop whereby red presents a counterclockwise and blue 
a clockwise circulation. The maximal current is $J_{\rm max} = 0.15\, et/\hbar$ for
{\bf a} and $0.13\, et/\hbar$ for {\bf b}. The current distribution of each of the two states has 
strong spatial variations and does not fulfill the continuity condition which, however, is restored 
for the total current.}
\label{Fig4}
\end{figure}

The numerical solution of the BdG equations with a self-consistency condition 
for the order parameter is adequate for $\simeq$~15 nm rings. However, to 
examine systems of micrometer size, the nodal states have to be described 
using a continuous gapless density of states. The flux induces a Doppler shift 
which modifies the states and alters their occupation near $E_F$, thereby 
causing an $h/e$ component of the current $J$. While the $h/2e$ component of 
$J\propto 1/d$, the $h/e$ component decreases with $1/d^2$ (see 
Appendix B). In quantitative agreement the $h/e$ component which, 
as compared to a ring of the size shown in Fig.~1, reduces by a factor of 60 
for a corresponding ring of 1~$\mu$m size, measured by the weight of its 
Fourier peak. Using typical parameter values for a YBCO ring of 1~$\mu$m size, 
the ratio of the $h/e$ versus the $h/2e$ component remains in the percent 
range. The frame width $w$ of the ring has little influence on the weight of 
the $h/e$ component for the loops with $w$ smaller than the penetration depth 
$\lambda$. A similar behavior is also shown by loops with $w>\lambda$, because 
only states that result in the current-transport channels within $\lambda$ 
affect significantly the $h/e$ component. 

Our calculations show that while changes in geometry, the number of 
transverse channels and elastic scattering by impurities modify the $J(\Phi)$ 
characteristics in detail, they do not eliminate the $h/e$ component. As long 
as the single particle states are well defined, also electronic correlation 
effects, which are responsible for the renormalization of states and of
coupling parameters, are not expected to bear a strong influence on the 
discussed phenomena.

The robust, magnetic-flux induced presence of currents that flow opposite to 
the main screening currents affect many properties of unconventional 
superconductor. Of particular importance are a resulting enhancement of the 
London penetration depth and a weakening of the rf-shielding. The 
$h/e$ periodicity of the supercurrent is a fundamental
property of loops formed by unconventional superconductors.

\newpage
\appendix


\section{Numerical Method}
To investigate ring geometries with finite width, we self-consistently solve the Bogoliubov - de Gennes (BdG) equations on the square frame shown in Fig.~\ref{Fig3} for the Hamiltonian 
\beq
{\cal H}=-t\sum_{\langle ij\rangle s}e^{i\varphi_{ij}}c_{is}^\dag c_{js}+\sum_{\langle ij\rangle}\left[\Delta_{ij}^* c_{j\downarrow}c_{i\uparrow}+\Delta_{ij} c^\dag_{i\uparrow}c^\dag_{j\downarrow}\right],
\label{s021}
\eeq
where
\beq
\Delta_{ij}=\frac{V}{2}\left(\langle c_{j\downarrow}c_{i\uparrow}\rangle-\langle c_{j\uparrow}c_{i\downarrow}\rangle\right)
\label{s082}
\eeq
is the order parameter defined on the two neighboring lattice sites $i$ and $j$. The pairing interaction strength is $V$ and appropriate phases for $d$-wave pairing: $\Delta_{i,i+\hat x}=-\Delta_{i,i+\hat y}$ are implicitly incorporated. A magnetic flux is represented by the Peierls phase factor $\varphi_{ij}=\frac{2\pi e}{h}\int_i^j{\bf A}({\bf r})\cdot {\rm d}{\bf r}$. We choose a vector potential of the form ${\bf A}=\phi(y,-x)/(2\pi r^2)$, yielding a flux threading the hole with no magnetic field penetrating the superconductor, where $\phi=\Phi\;e/h$ measures the flux in units of $h/e$.
\begin{figure}[h!]
\centering
\includegraphics[width=8cm]{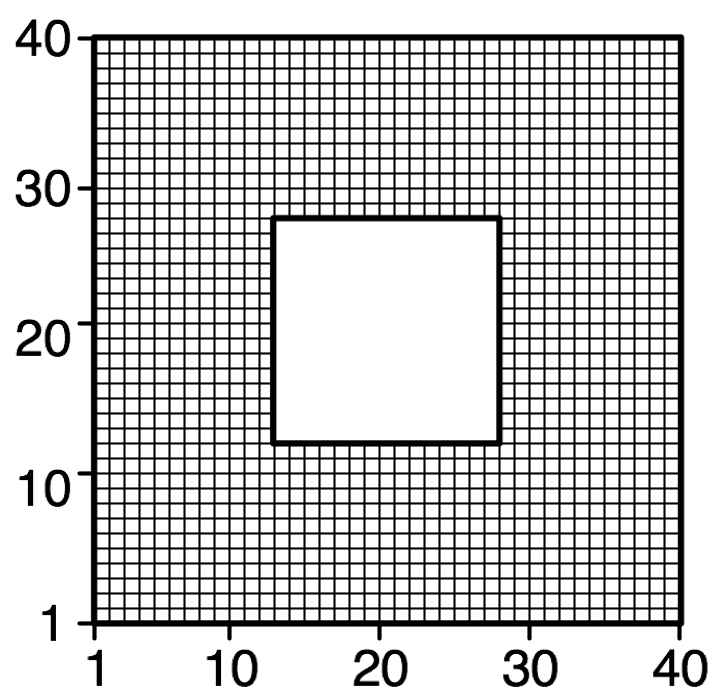}
\caption{\setlength{\baselineskip}{5mm}
A two dimensional discrete lattice for a square frame with open boundary conditions. The figure shows the standard geometry for a system size of 40$\times$40 lattice sites with a centered hole of 14$\times$14 lattice sites, on which calculations were typically performed.}
\label{Fig3}
\end{figure}
The Hamilton operator (\ref{s021}) is diagonalized by the Bogoliubov transformation
\bea
c_{i\uparrow}&=&\sum_n\left[u_{ni}\gamma_{n\uparrow}-v^*_{ni}\gamma_{n\downarrow}^\dag\right],
\label{s35}\\
c_{i\downarrow}&=&\sum_n\left[u_{ni}\gamma_{n\downarrow}+v^*_{ni}\gamma_{n\uparrow}^\dag\right].
\label{s36}
\eea
where $\gamma_{ns}^\dag$ and $\gamma_{ns}$ are creation and annihilation operators for fermionic Bogoliubov quasiparticles. The coefficients $u_n$ and $v_n$ have to fulfill the equation
\beq
\begin{pmatrix}H_0&\Delta\cr\Delta^\dag &-H_0^{\dag}\end{pmatrix}\begin{pmatrix}{u}_n\cr{v}_n\end{pmatrix}=E_n\begin{pmatrix}{u}_n\cr{v}_n\end{pmatrix}.
\label{s37}
\eeq
where the operators $H_0$ and $\Delta$ act on the ``single particle'' wave functions $u_n$ and $v_n$ as
\bea
H_0u_{ni}&=&-t\sum_{j}e^{i\varphi_{ij}}u_{nj}-\mu u_{ni},
\label{s38}\\
\Delta v_{ni}&=&\sum_j\Delta_{ij}v_{nj},
\label{s39}
\eea
and $\sum_j$ denotes the sum over all nearest neighbor sites of $i$. The order parameter $\Delta_{ij}$ is calculated self-consistently from
\beq
\Delta_{ij}=\frac{V}{2}\sum_n\left[u_{ni}v^*_{nj}+u_{nj}v^*_{ni}\right]\tanh\left(\frac{E_n}{2k_BT}\right),
\label{s41}
\eeq
where the sum runs over the positive eigenvalues $E_n$ only and $T$ is the temperature.  The current density $J_{ij}$ from lattice site $i$ to $j$ is
\bea
J_{ij}&=&-i\frac{et}{\hbar}\sum_s\left(c_{is}^\dag c_{js}e^{i\varphi_{ij}}-c_{sj}^\dag c_{is}e^{i\varphi_{ji}}\right)\\
&=&-4\frac{et}{\hbar}\sum_n\mbox{Im}\left[\left(u_{nj}u_{ni}^*f(E_n)+v_{nj}^*v_{ni}(1-f(E_n))\right)e^{i\varphi_{ij}}\right];
\label{s42}
\eea
$f(E)=1/(1+e^{E/k_BT})$ is the Fermi distribution function. The self-consistent solutions of the BdG equations on the square loop are characterized by the winding number $q$ of the phase of the order paraneter $\Delta_{ij}$ around the loop. For a fixed value of flux $\phi$, ground-state solutions in different $q$-sectors are found by choosing suitable starting values for the iterations in the self consistency loop.
\section{Multi-Channel Model for Large \boldmath$D$-Wave Rings}
Since the numerical method outlined above is not suited for calculations on loops of larger size say in the $\mu$m range, we use a multi-channel ring model which allows for an analytic calculation. A superconducting ring is thereby composed from many one dimensional (1$D$) loops with different radii. Each loop represents one current channel; the properties of the ring are obtained by integrating over its thickness. In 1$D$, the only spin-singlett pairing symmetry possible is $s$-wave pairing. To obtain the current characteristics of a $d$-wave loop, we use the following sceme: We obtain the supercurrent in the loop through an energy integration over the current contribution of all occupied eigenstates of the BCS Hamiltonian. In a circular loop (with no hybridization), the Doppler shift of the eigerenergies is a linear function of the flux independent of the pairing symmetry. The only way in which the symmetry influences the supercurrent is through its characteristic density of states (DOS). We therefore perform the following calculations for a 1$D$ $s$-wave loop and obtain the $d$-wave supercurrent for a quasi 1$D$ channel by inserting the (Doppler shifted) $d$-wave DOS (Fig.~\ref{Fig1}) into the final energy integration for the supercurrent (Eq.~(\ref{s29})).
\subsection{Superconducting State in a 1\boldmath$D$ \boldmath$s$-wave loop }
We describe the kinetic energy of the electrons on an individual flux threaded ring with $N$ discrete lattice sites
\beq
{\cal H}_0=\sum_{ks}\epsilon_{k-\phi}c_{ks}^\dag c_{ks},
\label{s04}
\eeq
by the tight binding dispersion
\beq
\epsilon_{k-\phi}=-2t\cos\left(\frac{k-\phi}{R}\right).
\label{s05}
\eeq
$\epsilon_{k-\phi}$ is the energy of a single particle state with angular momentum $\hbar k$ with $k\in\mathbb Z$. The radius of the ring measured in units of the lattice constant $a$ is $R=N/2\pi$. The BCS  pairing Hamiltonian has the form
\beq
{\cal H}={\cal H}_0+\sum_{k,q}\left[\Delta^*_k(q)c_{-k+q\downarrow}c_{k\uparrow}+\Delta_k(q)c^\dag_{k\uparrow}c^\dag_{-k+q\downarrow}\right],
\label{s09}
\eeq
where 
\beq
\Delta_k(q)= \sum_k\frac{V_{kk'}}{2}\left[\langle c_{k'\downarrow}c_{-k'+q\uparrow}\rangle-\langle c_{k'\uparrow}c_{-k'+q\downarrow}\rangle\right]
\label{s010}
\eeq
is the superconducting order parameter and $q\in\mathbb Z$ its winding number. For a perfectly circular ring geometry, the winding number can be identified with the angular momentum $\hbar q$ of a Cooper pair. Choosing the pairing energy $V_{kk'}=V$ independent of $k$ and $k'$ leads to pairing in the $s$-wave channel, which is the only possibility in one space dimension.
Since we are interested in low temperature properties, we assume that the superconducting condensate in its ground state is characterized by the quantum number $q(\phi)$, which changes its value as a function of flux $\phi$  whenever the total energies for two different $q$-values become degenerate. Up to finite size effects, the $q$-number of the ground state changes to the next integer whenever $\phi$ crosses the flux values $(2n-1)/4$, $n\in\mathbb Z$:
\beq
q(\phi)=\mbox{int}\left(2\phi+\frac{\mbox{sign}(\phi)}{2}\right),
\label{s8}
\eeq
where for positive (negative) $x$, $\mbox{int}(x)$ is the largest (smallest) integer number equal or smaller (larger) than $x$.
We therefore choose an ansatz for $\Delta_k(q)$ of the form
\beq
\Delta_k(q')=\delta_{q(\phi),q'}\Delta_{k},\label{s7}
\eeq
with arbitrary $k$-dependence of $\Delta_{k}$. For $s$-wave pairing $\Delta_{k}=\Delta$ is constant and $\Delta_k(q)=\Delta(q)$. 
With this ansatz the diagonalization of the Hamiltonian (\ref{s09}) leads to the energy spectrum 
\beq
E_{\pm}(k,\phi)=\frac{\epsilon_{k-\phi}-\epsilon_{-k-\phi+q}}{2}\pm\sqrt{\Delta^2+\left(\frac{\epsilon_{k-\phi}+\epsilon_{-k-\phi+q}}{2}\right)^2}.\label{s10}
\eeq
The energies $E_\pm(k,\phi)$ shift with flux and for $\phi\neq n$ the particle-hole symmetry of the spectrum is broken. Near the Fermi energy, in different $q$-sectors, the Doppler shift $e(\phi)$ of the eigenenergies is found by expanding $E_\pm(k,\phi)$ in $\phi$, leading to $e(\phi)=\pm(\phi-q/2)\:2t/R+{\cal O}((\phi/R)^2)$. If therefore $\Delta(q)\leq e(\phi=1/4)=t/2R$, where the condensate changes $q$ form $0$ to $1$, the indirect energy gap closes and the occupation of states changes. For $\phi>1/4$, the pairing of electrons in states with total angular momentum $q \neq0$ according to Eq.~(\ref{s010}) becomes favorable.
In one dimension, the gap closes exactly at the depairing velocity of the condensate beyond which no self-consistent solution of the order parameter exists \cite{Bagwell}.
\subsection{Current in a \boldmath$d$-wave loop}
In this section, we first derive a expression in form of an energy integration for the supercurrent in a $s$-wave loop which is then transformed into a $d$-wave loop as described above.

In the nearest-neighbor tight binding model for a 1$D$ ring, the current is given by
\beq
J(R)=\frac{e}{h}\sum_{ks}J_k(R)n_{ks}=\frac{e}{\hbar R}\sum_{ks}\frac{\partial\epsilon_{k-\phi}}{\partial k}\langle c_{ks}^\dag c_{ks}\rangle,\label{s13}
\eeq
where $J_k={\partial\epsilon_{k-\phi}}/{\partial k}$ is the group velocity of the state with angular momentum $\hbar k$ and the occupation probability $n_{ks}$ is obtained using a Bogoliubov transformation:
\beq
n_{ks}=\langle c_{ks}^\dag c_{ks}\rangle=\sum_{\alpha=\pm1}\frac{\alpha}{2}\left(\frac{\epsilon_{k-\phi}+\epsilon_{-k-\phi+q}}{\sqrt{4\Delta^2+\left(\epsilon_{k-\phi}+\epsilon_{-k-\phi+q}\right)^2}}+\alpha\right)f(E_\alpha(k,\phi)).
\label{s14}
\eeq
Eqs.~(\ref{s13}) and (\ref{s14}) are a closed-form solution for the total current in a superconducting flux threaded ring. The sum over $k$ has to be computed numerically, though. As shown  below, the expansion of $J_k(R)$ in powers of $\phi/R$, provides a $\phi$ independent contribution which is paramagnetic for $q=0$ and diamagnetic for $q=1$ plus a contribution linear in $\phi$, which is diamagnetic for $q=0$ (Meissner effect) and paramagnetic for $q=1$ \cite{Scalapino}.

\begin{figure}[h!]
\centering
\vskip2mm
\includegraphics[width=8cm]{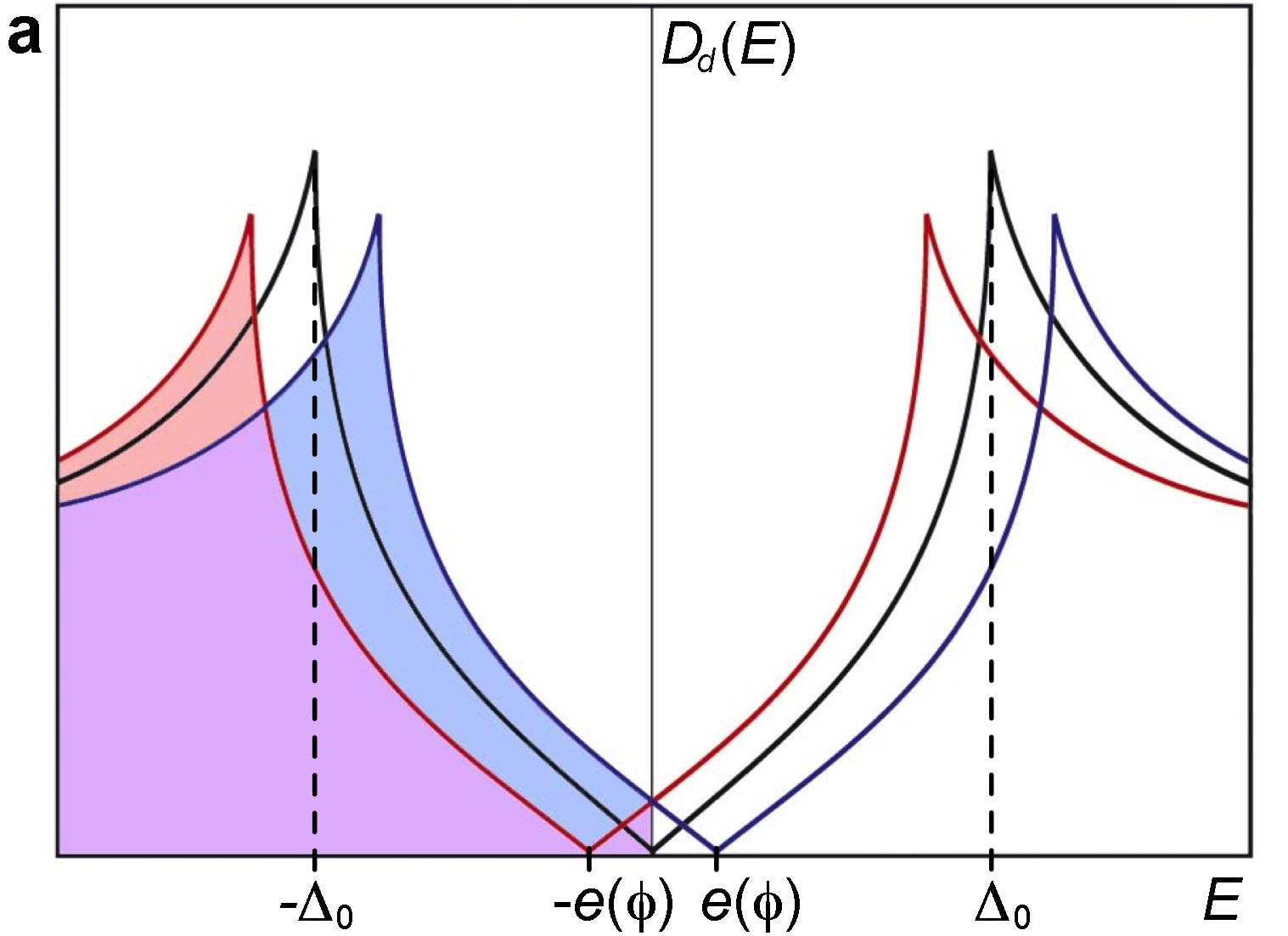}\\[1mm]
\includegraphics[width=8cm]{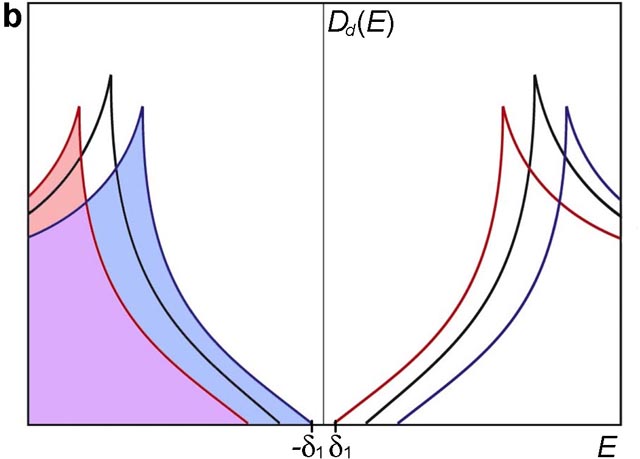}
\caption{\setlength{\baselineskip}{5mm}
Scheme for the density of states of a $d$-wave superconductor for a finite flux $|\phi|<1/4$ ($q=0$) ({\bf a}) and for $1/4<\phi<3/4$ ($q=1$) ({\bf b}). The energies are Doppler shifted to higher (red) or lower energies (blue). This results in a double-peak structure of the coherence peaks and for $q=0$ in an overlap of the upper and lower band in the region $-e(\phi)<E<e(\phi)$ $^($\cite{Maki}$^)$. States in the upper band become partially occupied. For $q=1$ there is a gap of width $2\delta_1\equiv2\delta_1(\phi)$. The black line in {\bf b} represents the density of states for $\phi=1/2$.}
\label{Fig1}
\end{figure}

In the following we restrict the discussion to the flux interval $-1/4\leq\phi\leq1/4$ where $q=0$ in the ground state. 
We assume that $R\gg1$ and expand $\epsilon_{\pm k-\phi}$ and $J_{k}$ in $\phi/R$ for $k\geq0$ as:
\bea
\epsilon_\pm=\epsilon_{\pm k-\phi}&=&\epsilon_{k}\pm\frac{2t}{R}\phi\sqrt{1-\left(\frac{\epsilon_k}{2t}\right)^2}+{\cal O}((\phi/R)^2),
\label{s23}\\[0.2cm]
J_\pm(R)=J_{\pm k}(R)
&=&\mp\frac{2te}{R}\left[\sqrt{1-\left(\frac{\epsilon_k}{2t}\right)^2}\mp\frac{\phi}{R}\frac{\epsilon_k}{2t}\right]+{\cal O}((\phi/R)^2).
\label{s242}
\eea
To leading order, the quasiparticle energies in the superconducting state become
\beq
E_\pm=E_\pm(\pm k,\phi)=E\pm\frac{2t}{R}\phi\sqrt{1-\left(\frac{\epsilon_k}{2t}\right)^2}+{\cal O}((\phi/R)^2),
\label{s243}
\eeq
with $E=\sqrt{\epsilon_k^2+\Delta^2}$ if $\epsilon_k >0$ and $E=-\sqrt{\epsilon_k^2+\Delta^2}$ if $\epsilon_k <0$. In the vicinity of the Fermi energy $E_F$, this simplifies to $\epsilon_\pm=\epsilon_k\pm e(\phi)$ and $E_\pm=E\pm e(\phi)$.
Converting the sum over $k$ in Eq.~(\ref{s13}) to an integral over the normal state energy $\epsilon_k$, the total current becomes
\beq
J(R)=\int {\rm d}\epsilon\:D(\epsilon)\:n_+(\epsilon)J_{+}(R,\epsilon)+\int {\rm d}\epsilon\:D(\epsilon)\:n_-(\epsilon)J_{-}(R,\epsilon)+{\cal O}((\phi/R)^2),
\label{s28}
\eeq
where
\beq
n_\pm(\epsilon)= n_{\pm k}=\sum_{\alpha=\pm1}\frac{\alpha}{2}\left(\frac{|\epsilon|}{\sqrt{\Delta^2+\epsilon^2}}+\alpha\right)f(E_\pm)
\label{s141}
\eeq
and $D(\epsilon)$ is the DOS of the normal state. With $\epsilon=\pm\sqrt{E^2-\Delta^2}$ we rewrite Eq.~(\ref{s28}) as
\bea
J(R)&=&N_0\int_{-2t}^{2t}dE\Bigg[\sum_{\alpha=\pm1}\frac{1}{2}\left(\alpha+D_s(E)\right)f(E+\delta(\phi))J_-(E)\nonumber\\
&&\makebox[15mm]{}+\sum_{\alpha=\pm1}\frac{1}{2}\left(\alpha+D_s(E)\right)f(E-\delta(\phi))J_+(E)\Bigg]
\label{s29}
\eea
where we assume $D(\epsilon)=N_0$ constant in the vicinity of $E_F$ and $D_s(E)$ is the DOS in the $s$-wave superconductor: $D_s(E)=|E|/\sqrt{E^2-\Delta^2}$ if $|E|>\Delta$ and $D_s(E)=0$ if $|E|<\Delta$. At $T=0$ the current can be separated into two contributions $J(R)=J_1(R)+J_2(R)$, where $J_1$ contains all the contributions from the states which are below $E_F$ within the interval $-1/4<\phi<1/4$ representing the standard supercurrent and $J_2$ contains the additional contributions from $J(R)$ which appear if $e(\phi)>\Delta$. One finds:
\bea
\label{s30}
J_1(R)&=&-N_0\frac{2e}{hR^2}\phi\int_{-2t}^{-e(\phi)}dE|E| \cong-N_0\frac{e}{h}\phi\left(\frac{2t}{R}\right)^2,\\[0.5cm]
J_2(R)&=&N_0\frac{2te}{hR}\int_{-e(\phi)}^{e(\phi)}dED_s(E)+{\cal O}\left(\frac{\phi^3}{R^5}\right),
\label{s31}
\eea
where the upper integration boundary in Eq.~(\ref{s30}) is extended to zero. 

\begin{figure}[h!]
\centering
\includegraphics[width=8cm]{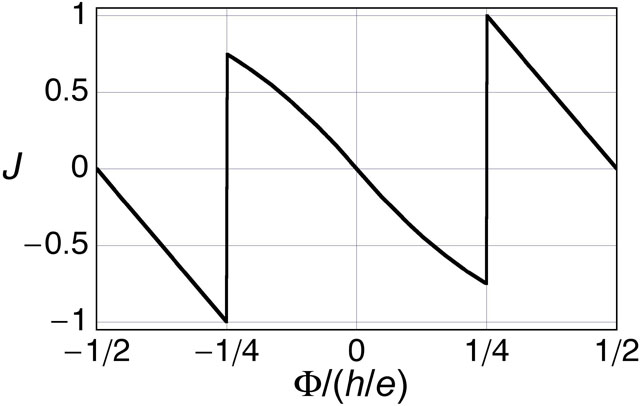}
\caption{\setlength{\baselineskip}{5mm}
The current in a thin $d$-wave ring as a function of flux $\Phi$ (in arbitrary units). Shown is the result of the multi-channel model 
for the characteristic value $2t/(\Delta_0R)=0.4$. For $-h/4e < \Phi<h/4e$, where $q=0$, the current is reduced by a contribution proportional to $\Phi^2$, whereas it is strictly linear in $\Phi$ otherwise. This gives rise to an overall current periodicity of $h/e$.}
\label{Fig2}
\end{figure}

We replace now $D_s(E)$ in Eq.~(26) by the DOS $D_d(E)$ of a $d$-wave superconductor as shown in Fig.~\ref{Fig1}~{\bf a}. For finite flux $\phi$, all energy levels are shifted according to the magnetic moment of their current; this results in a Doppler shift of the coherence peaks \cite{Maki}. In the relevant regime $\Delta_0 > e(\phi)$, it is sufficient to approximate $D_d(E)\simeq|E|/\Delta_0$ (Fig.~\ref{Fig1}~{\bf a}) and
\beq
J_2(R)\simeq\frac{N_0}{\Delta_0}\frac{2te}{hR}\int_{-e(\phi)}^{e(\phi)}dE\:|E|=\frac{N_0}{\Delta_0}\frac{e}{h}\phi^2\left(\frac{2t}{R}\right)^3.
\label{s33}
\eeq
The total current $J(R)=J_1(R)+J_2(R)$ of this channel becomes
\beq
J(R)=-N_0\frac{e}{h}\phi\left(\frac{2t}{R}\right)^2\left[1-\phi\frac{2t}{\Delta_0R}\right].
\label{s34}
\eeq
The normal state DOS, $N_0=R/(2t)$, is itself a function of $R$. The total current $J$ for $q=0$ in a ring of finite thickness $D$ and inner radius $R_<$ is obtained from
\bea
J&=&\int_{R_<}^{R_<+D}dR(J_1(R)+J_2(R)).
\label{s35}\\
&=&-N_0\frac{e}{h}\phi(2t)^2\frac{d}{R_<(R_<+D)}\left[1-\phi\frac{2t}{\Delta_0}\frac{D+2R_<}{2R_<(R_<+D)}\right].
\label{s36}
\eea
In the limit of thin rings ($R_<\gg D$), we introduce $d=2R_<$ in units of the lattice constant $a$ and find that the ratio
\beq
\frac{J_2}{J_1}=2\frac{2t}{\Delta_0}\frac{\phi}{d}
\label{s45}
\eeq
shows the same power law in $1/{d}$ as for a single channel.

For $q=\pm1$, an energy gap $\Delta(q)>e(\phi)$ persists for all $\phi$, thus a flux induced effective gap $2\delta_1(\phi)$ is present, as shown in Fig.~\ref{Fig1}~{\bf b}. Therefore calculations as above are valid also for this flux window, however $J_2(R)=0$. Only the standard supercurrent component  contributes to the total current with $J(R)=J_1(R)$ for each channel (Fig.~\ref{Fig2}). 

Because $J_2$ is finite for even $q$ but zero for odd $q$, whereas $J_1$ is identical for all $q$, we find that $J_1$ is periodic with $h/2e$ and $J_2$ with $h/e$. The result in Eq.~(\ref{s45}) implies that the ratio of the $h/e$ and the $h/2e$ Fourier component of the total current scales with the inverse ring diameter.

\clearpage

\noindent
\textbf{References}
{}

\vspace{5mm}

\noindent

\vspace{5mm}

\noindent
{\bf Acknowledgements} The authors gratefully acknowledge helpful discussions with Dieter Vollhardt. This work was supported by the DFG (SFB 484), the EC (Nanoxide), and the ESF (THIOX).

\end{document}